# Magnetic properties of molecular beam epitaxy-grown ultrathin $Cr_2Ge_2Te_6$ films down to monolayer limit on Si substrates

Pengfei Ji[1,#], Ruixuan Liu[1,#], Tianchen Zhu[1,#], Jinxuan Liang[1], Yang Chen[1], Yitian Tong[1], Yunhe Bai[1], Zuhan Geng[1], Fangting Chen[2], Yunyi Zang[3], Xiyu Hong[1], Jiatong Zhang[1], Luyi Yang[1], Qi-Kun Xue[1,3,4], Ke He[1,3,5,6*], and Xiao Feng[1,3,5,6*]

[1]State Key Laboratory of Low Dimensional Quantum Physics, Department of Physics, Tsinghua University, Beijing 100084, China

[2]Shanghai Aerospace Electronic Communication Equipment Research Institute, Shanghai 201108, China

[3]Beijing Academy of Quantum Information Sciences, Beijing 100193, China

[4]Southern University of Science and Technology, Shenzhen 518055, China

[5]Frontier Science Center for Quantum Information, Beijing 100084, China

[6]Hefei National Laboratory, Hefei 230088, China

[#]These authors contribute equally to this work

[*]Corresponding authors: E-mail: kehe@mail.tsinghua.edu.cn; xiaofeng@mail.tsinghua.edu.cn

**$Cr_2Ge_2Te_6$, a prototypical van der Waals ferromagnetic semiconductor, have attracted significant interest for its potential applications in high-performance spintronics. However, the magnetic ground state of monolayer $Cr_2Ge_2Te_6$ remains elusive due to fragile and irregular-shaped thin flake samples with weak magnetic signals. Here, we successfully grow uniform ferromagnetic $Cr_2Ge_2Te_6$ films down to monolayer by molecular beam epitaxy. By exploiting a self-limiting growth mode, we achieve synthesis of uniform monolayer $Cr_2Ge_2Te_6$ films across entire millimeter-scale Si substrates. Through a combination of superconducting quantum interference device magnetometry and anomalous Hall effect measurements, we establish that monolayer $Cr_2Ge_2Te_6$ exhibits intrinsic ferromagnetism with perpendicular magnetic anisotropy below ~ 10 K, albeit with strong magnetic fluctuations characteristic of its two-dimensional nature.**

**Furthermore, a systematic thickness-dependent study reveals a crossover from this fluctuation-dominated two-dimensional magnetism turns into conventional long-range ferromagnetic order as the film thickness increases. Our work not only definitively establishes the intrinsic ferromagnetic ground state of monolayer $Cr_2Ge_2Te_6$, but also provides a scalable, silicon-compatible route for preparing the two-dimensional magnet for future spintronic or quantum devices.**

**Introduction**

Van der Waals (vdW) magnetic materials have emerged as a focal point of intensive research in recent years. The weak interlayer bonding facilitates mechanical exfoliation and the atomically thin flakes provide an ideal platform for probing magnetic properties and mechanisms in the two-dimensional (2D) limit.[1-3] $Cr_2Ge_2Te_6$, as an archetypal vdW ferromagnetic semiconductor with bandgap of 200 meV,[4-9] is particularly attractive for high-performance spintronics [10-14] and novel quantum phases such as quantum anomalous Hall system[15-24] and topological superconductor.[25] Although the Curie temperature ($T_C$) of bulk $Cr_2Ge_2Te_6$ is ~ 65 K, it can be substantially enhanced by impurity doping or tensile strain.[26-35] Ferromagnetic order has been observed in $Cr_2Ge_2Te_6$ thin flakes down to bilayer, which has stimulated extensive investigations into the vdW magnet[36-39] and various heterostructures based on it.[40-45]

Despite substantial experimental efforts, the magnetism of monolayer (1 L) $Cr_2Ge_2Te_6$ remains a mystery.[46-50] Well established ferromagnetism has yet to be observed in 1 L $Cr_2Ge_2Te_6$, and it is not clear whether it represents its intrinsic property or merely stems from sample degradation. Two primary challenges impede the investigation of intrinsic magnetism of 1 L $Cr_2Ge_2Te_6$. First, $Cr_2Ge_2Te_6$ is sensitive to surface contamination, particularly in the 1 L limit where the surface-to-volume ratio is maximized. Conventional glovebox environment used for mechanical exfoliation may not provide sufficient protection to prevent degradation of 1 L samples.[51,52] Second, the small size and irregular shape of exfoliated $Cr_2Ge_2Te_6$ thin flakes pose severe difficulties in direct magnetization measurements such as superconducting quantum interference device (SQUID) magnetometer. Furthermore, the fragility and irregularity of Cr$_2$Ge$_2$Te$_6$ thin flakes also present significant obstacles to fabrication of Cr$_2$Ge$_2$Te$_6$-based vdW heterostructures for advanced studies and applications.

These challenges can be effectively addressed in molecular beam epitaxy (MBE)-grown Cr$_2$Ge$_2$Te$_6$ thin films. MBE is an ultra-high vacuum (UHV) thin film deposition technique that enables highly controlled purity and atomic-layer-level thickness in macroscopic size. MBE growth of Cr$_2$Ge$_2$Te$_6$ thin films down to 6 nm have been achieved on (Bi,Sb)$_2$Te$_3$ thin film-covered InP(111). However, the small bulk gap (~ 200 meV) and gapless surface states of (Bi,Sb)$_2$Te$_3$ as a topological insulator, as well as inevitable thickness variation in MBE-grown thin films could complicate the investigations in few-layer Cr$_2$Ge$_2$Te$_6$ films. In this work, we achieved MBE growth of ultra-thin Cr$_2$Ge$_2$Te$_6$ films down to 1 L on Si substrates. The growth of 1 L Cr$_2$Ge$_2$Te$_6$ films was found to be self-limiting, resulting a uniform 1 L-thick Cr$_2$Ge$_2$Te$_6$ films across the whole macroscopic-sized substrate. The well-controlled growth condition enables a systematic investigation of the magnetic evolution of the prototypical vdW ferromagnet with single-layer precision.[39,53,54]

## Results

High quality Cr$_2$Ge$_2$Te$_6$ thin films were grown on Si(111)-(7×7) substrates via co-evaporation of Cr, Ge and Te elemental sources (see Methods for the details). The reflective high energy diffraction (RHEED) pattern of the MBE-grown Cr$_2$Ge$_2$Te$_6$ films displays sharp diffraction streaks, indicating the high-quality flat surface (Fig. S1a). The in-plane lattice constant of Cr$_2$Ge$_2$Te$_6$ is approximately 2.3% larger than that of Si(111)-√3×√3$R$30°, constituting a nearly lattice-matched interface, as shown in Fig. 1a. The x-ray diffraction (XRD) pattern exhibits sharp diffraction peaks of Cr$_2$Ge$_2$Te$_6$ and Si, with no evidence of other phases (Fig. S1b).

Figure 1b presents an atomic-resolution cross-sectional scanning transmission electron microscopy (STEM) image of a 35 L Cr$_2$Ge$_2$Te$_6$ film grown on Si(111). A zoom-in image around the interface is shown in Fig. 1c. The atomic structures of both Si and Cr$_2$Ge$_2$Te$_6$ film are clearly resolved, with atomically sharp interface. In

$Cr_2Ge_2Te_6$ film, the heaviest Te atoms exhibit the highest contrast in STEM images. At the interface, the Si lattice is terminated by a monolayer of atoms displaying significantly higher contrast than Si atoms, though comparable to Te atoms in the $Cr_2Ge_2Te_6$ film. This observation suggests that at the initial stage of $Cr_2Ge_2Te_6$ growth, Si surface is first passivated by a monolayer of Te atoms, followed by growth of $Cr_2Ge_2Te_6$ film analogous to van der Waals epitaxy.[55] This interpretation is supported by the observation that the interlayer spacing between the Te passivation layer and the subsequent $Cr_2Ge_2Te_6$ layer closely matches the van der Waals gap between adjacent $Cr_2Ge_2Te_6$ layers.

Fourier analysis of the high-resolution transmission electron microscopy (HRTEM) image was performed to determine the in-plane lattice constants of $Cr_2Ge_2Te_6$ across different layers, as shown in the right panel of Fig. 1b. The in-plane lattice constant of the first two $Cr_2Ge_2Te_6$ layers near to the interface is ~ 6.9 Å (the lattice constant is twice of the atomic column spacing along the direction) which is larger than that of Si(111)-√3×√3$R$30° (~ 6.651 Å). For layers beyond the initial two, the lattice constant increases to ~ 7.0 Å, consistent with that of $Cr_2Ge_2Te_6$ films grown on $(Bi,Sb)_2Te_3$ substrates.[56-58]

Figure 1d presents the perpendicular magnetic field dependence of magnetoresistance (MR) curves of a 25 nm $Cr_2Ge_2Te_6$ film measured at various temperatures. Below 60 K, butterfly-shaped MR curves emerge, suggesting the ferromagnetic orders in the film. Due to the high resistivity (Fig. S1d), reliable anomalous Hall effect (AHE) data cannot be obtained at lower temperatures. Magnetic properties were further characterized using a SQUID magnetometer. Figure 1e displays the magnetic moment ($m$) vs. $\mu_0 H_\perp$ curves (*M-H* curves) of the same 25 nm $Cr_2Ge_2Te_6$ film at different temperatures. Rectangular hysteresis loops are observed from 2 K to 60 K, indicating ferromagnetism with perpendicular magnetic anisotropy, consistent with the MR curves. The magnetic properties of the 25 nm $Cr_2Ge_2Te_6$ film grown on Si are similar to those on $(Bi,Sb)_2Te_3$-covered InP.[56-58]

Investigating the magnetic properties of few-layer $Cr_2Ge_2Te_6$ films presents bigger challenges. The fragile samples were protected by an in-situ grown capping layer in

MBE chamber. To distinguish the weak magnetic signals of the films from those of possible contaminations, magnetic characterizations should be performed using both SQUID and AHE. Only consistent results through both techniques can be attributed to the intrinsic sample properties. The highly insulating few-layer $Cr_2Ge_2Te_6$ at low temperature preludes direct Hall measurements, an adjacent non-magnetic conducting layer was employed for indirect probing.[60,61] Figures. 1f and 1h display the hysteresis loops of a 6 L $Cr_2Ge_2Te_6$ with a 4 nm $Bi_2Te_3$ capping layer measured at different temperatures by SQUID and AHE, respectively. The hysteresis loops measured with the two different methods at each temperature have nearly the same shape. Figures 1g and 1i display temperature dependences of normalized anomalous Hall resistivity (zero field Hall resistivity $\rho_{yx}$ normalized to $\rho_{yx}$ at 2 K and 2 T, extracted from Fig. 1f) and magnetic moment (extracted from Fig. 1h), respectively. Both show similar temperature dependences behaviors and the $T_C$s are both ~ 60 K. The rapid growth of normalized anomalous Hall resistivity below 10 K can be attributed to the enhancement of the anomalous Hall coefficient where magnetization tends to saturation (Fig. 1i).[59] The SQUID and AHE measurements with conducting capping layer show highly consistent results.

However, there could be unintentional Cr dopants diffused from the adjacent $Cr_2Ge_2Te_6$ layer which contribute the spurious magnetic signals. Different capping layers ($Bi_2Te_3$, $Sb_2Te_3$ and CdTe) were applied. The $Cr_2Ge_2Te_6$ films with different capping layers all exhibit atomically sharp interfaces in STEM images (Fig. S2). Figures 1j and 1k display *M-H* curves and *M-T* curve of a 6 L $Cr_2Ge_2Te_6$ film with CdTe capping layer (~ 4 nm) at various temperatures. The data are quantitatively consistent with those of the film with $Bi_2Te_3$ capping layer (Figs. 1h and 1i). Obviously, neither the capping layer materials nor the measurement methods influence the magnetic measurement results, which can only be attributed to the $Cr_2Ge_2Te_6$ films.

Figure 2a shows a large-scale STEM image of a 1 L $Cr_2Ge_2Te_6$ film grown on Si(111) with a $Bi_2Te_3$ capping layer. Figure 2b displays an atomic-resolution STEM image of a similar sample. A single $Cr_2Ge_2Te_6$ layer is clearly resolved between Te-terminated Si substrate and $Bi_2Te_3$ capping layer. Remarkably, across the entire areas of the STEM images (up to 150 nm, as shown in Fig. S3), only 1 L $Cr_2Ge_2Te_6$ is observed, with no trace of thicker or thinner regions. The atomic force microscope (AFM) image (3 μm × 3 μm) of a film grown under the same condition (Fig. 2c) reveals the typical step-terrace morphology of Si surface, with step heights corresponding to the Si(111) interlayer distance. Neither islands nor hole-like depressions of $Cr_2Ge_2Te_6$ are observed. These observations indicate the formation of a complete, uniformly 1 L $Cr_2Ge_2Te_6$ film across the entire Si substrate without detectable thickness variation, which is barely seen MBE in grown films. Unexpectedly, a $Cr_2Ge_2Te_6$ film remains 1 L even when the growth duration varies by ~ 20%, suggesting a self-limiting growth mode (Fig. S4), similar to that reported in epitaxial graphene on SiC.[62] Although the underlying mechanism needs further investigation, the self-limiting growth ensures formation of macroscopically uniform 1 L $Cr_2Ge_2Te_6$ which facilitates not only investigation on its intrinsic magnetic properties, but also scalable fabrication of spintronic devices based on 1 L $Cr_2Ge_2Te_6$. The self-limiting growth only occurs in 1 L $Cr_2Ge_2Te_6$. Thicker films exhibit unavoidable thickness variations which are typical in conventional MBE growth.

A close examination of the STEM image of Fig. 2b reveals fluctuations in both the position and the contrast of the atoms in 1 L $Cr_2Ge_2Te_6$ film, which is not observed in thicker samples. The fluctuations likely result from nanoscale ripples inherent to 2D nature of single-layer $Cr_2Ge_2Te_6$ films.[63] The observation of ripples suggests that the monolayer $Cr_2Ge_2Te_6$ film is close to a freestanding 2D system, only weakly bonded with Si substrate and $Bi_2Te_3$ capping layer.

The magnetic properties of 1 L $Cr_2Ge_2Te_6$ films with different capping layers were investigated with both AHE and SQUID measurements. In SQUID measurements, the samples as well as the sample rod and holders were very carefully handled to avoid the contaminations. Otherwise, the rather weak magnetic signals of 1 L $Cr_2Ge_2Te_6$ films (close to the measurement limit of SQUID ~ $2\times10^{-7}$ emu) would be buried in the background noise. The magnetic signal from the substrate and a linear background are subtracted from the raw data. The detailed data processing procedure is explained in Fig. S8.

Figures 2d-2k display the magnetic properties of 1 L $Cr_2Ge_2Te_6$ films with $Bi_2Te_3$(4 nm) (Figs. 2d-2f), $Sb_2Te_3$ (4 nm) (Figs. 2g-2i), and CdTe (4 nm) (Figs. 2j and 2k) capping layers measured by both AHE (except for the one with CdTe capping layer) and SQUID at 2 K. The films with different capping layers all exhibit hysteresis loops in both the normalized $\rho_{yx}$-H curves and M-H curves under perpendicular magnetic field ($H_\perp$). Meanwhile, M-H curves under in-plane field ($H_{//}$) show no hysteresis. The observation indicates the ferromagnetism with perpendicular magnetic anisotropy in all the samples. The measured magnetic moments at saturation magnetization (1 T) in different samples are also similar, $3\sim4\times10^{-7}$ emu. The estimated atomic moment is $\sim$ 1.35 $\mu_B$, smaller but at the same order of magnitude of $Cr^{3+}$ ion, which is reasonable considering the defects and stronger fluctuations in 1 L $Cr_2Ge_2Te_6$ samples.

Figures. 2l-2n show the temperature-dependent normalized $\rho_{yx}$-H curves and M-H curves of 1 L $Cr_2Ge_2Te_6$ film with $Bi_2Te_3$ capping layer. The normalized $\rho_{yx}$-H curves and M-H curves under $H_\perp$ show hysteresis below $\sim$ 5 K, while no hysteresis is observed in-plane M-H curves at any temperature. Clearly, the $T_C$s of 1 L $Cr_2Ge_2Te_6$ film measured by AHE and SQUID have no detectable difference. 1 L $Cr_2Ge_2Te_6$ films with other capping layers also exhibit similar $T_C$, as shown in Figs. S5-S7.

The magnetic properties independent of both capping layers and measurement methods are not likely contributed by magnetic contaminations nor by the capping layers with unintentional Cr dopants, but by the intrinsic properties of 1 L $Cr_2Ge_2Te_6$ films. Slightly difference in the coercive fields is observed in the films with different capping layers, probably resulting from the influence of capping layers on the magnetic anisotropy.

Figure 3 summarizes the systematic thickness and temperature dependence of the Hall effect in $Bi_2Te_3$-capped $Cr_2Ge_2Te_6$ films ranging from 1 L to 6 L, which demonstrate the evolution of magnetism from 2D limit to bulk-like by measuring the anomalous Hall effect (AHE). The Hall resistivity is normalized to the value at 2 T and at 2 K for each sample. At 2 K, the magnetic hysteresis loops are observed in all the films. As the film thickness increases, the loops shape evolves from canted to rectangular. The shape evolution is quantified by the thickness dependence of the remanence ratio $\rho_{yx}$(0 T)/$\rho_{yx}$(2 T) and $H_C$, shown in Fig. 4a, respectively. With increasing temperature, hysteresis loops in different samples diminishes and vanishes at different $T_C$s. The thickness dependence of $T_C$ is summarized in Fig. 4b. The remanence ratio $\rho_{yx}$(0 T)/$\rho_{yx}$(2 T), $H_C$ and $T_C$ show similar thickness dependent behavior. Below 3 L, they show an abrupt increase with increasing thickness, followed by a gradual increase to 6 L.

Figure 4c displays the temperature dependences of the normalized anomalous Hall resistivity $\rho_{yx}$ (0 T)/$\rho_{yx}$ (2 T, 2 K) of the films of different thicknesses. The abrupt increase in the curves at low-temperature (below 5 K) arises from enhancement of the anomalous Hall coefficients, as discussed earlier. Here, we focus on the higher-temperature behavior which primarily reflects the $M$-$T$ curve. Strikingly, the films of different thicknesses exhibit distinct $M$-$T$ behaviors. The 5 L and 6 L films display upward-convex curves below $T_C$: magnetization increases rapidly with decreasing temperature at first and then tends to saturation, analogous to typical $M$-$T$ curves in building of long-ranged ferromagnetic order. In contrast, the 1 L and 2 L films exhibit downward-convex curves: magnetization keeps growing with decreasing temperature without saturation, analogous to the $M$-$T$ curve of a paramagnet. The paramagnet-like $M$-$T$ curves indicate strong magnetic fluctuations that impede formation of long-range ferromagnetic order even below $T_C$. The intermediate 3 L and 4 L samples, representing a crossover regime, exhibit approximately linear $M$-$T$ behavior below $T_C$.

**Discussion**

1 L $Cr_2Ge_2Te_6$ includes only one atomic layer of magnetic atoms ($Cr^{3+}$), making it a model 2D magnet subject to strong magnetic fluctuations. According to the Mermin-Wagner theorem, long-range ferromagnetic order can only hold in 2D Ising magnets.[64,65] Consequently, $T_C$ of a 2D magnet is fundamentally limited by the magnetic anisotropic energy. The small perpendicular magnetic anisotropic energy of 1 L $Cr_2Ge_2Te_6$ leads to a low $T_C$. In bulk $Cr_2Ge_2Te_6$, interlayer exchange coupling suppresses magnetic fluctuations, and $T_C$ is governed by the exchange couplings, particularly the relatively weak interlayer coupling. In a 2 L $Cr_2Ge_2Te_6$ film, each $Cr_2Ge_2Te_6$ layer couples to one adjacent $Cr_2Ge_2Te_6$ layer, half of the bulk case where each layer has two nearest-neighbor layers. The experimentally observed $T_C$ of 2 L $Cr_2Ge_2Te_6$ film (~ 35 K) is roughly half of the bulk value (~ 65 K). As thickness increases, the average number of the nearest-neighbor layers increases, leading to an enhanced $T_C$. Using the model from Ref. 1, we calculate $T_C$s of $Cr_2Ge_2Te_6$ films of different thicknesses, as plotted in Fig. 4b (red data points), which are well consistent with the measured data. Application of a perpendicular magnetic field could suppress the magnetic fluctuations and is thus expected to enhance the $T_C$ in a 1 L $Cr_2Ge_2Te_6$ film. Figure 4d displays temperature-dependent normalized anomalous Hall resistivity of 1 L (blue) and 6 L (red) $Cr_2Ge_2Te_6$ films at 0 T, 0.1 T, 0.3 T, and 1 T, respectively. With increasing magnetic field, $T_C$ of 1 L $Cr_2Ge_2Te_6$ film is significantly enhanced from 10 K at 0 T to 40 K at 0.1 T. In contrast, $T_C$ enhancement under magnetic field is less significant for 6 L $Cr_2Ge_2Te_6$ film.

## Conclusion

Due to self-limiting growth mode of $Cr_2Ge_2Te_6$ on silicon substrates, we have successfully grown $Cr_2Ge_2Te_6$ thin films that remains monolayer across the entire macroscopic-sized Si substrates and are well protected by various capping layers. By combining measurements of AHE and SQUID magnetometry and assessing the impact of different capping layers, we directly address the fundamental question of whether single $Cr_2Ge_2Te_6$ possesses intrinsic ferromagnetic order. 1 L $Cr_2Ge_2Te_6$ possesses ferromagnetic order below 10 K with perpendicular magnetic anisotropy, as evidenced by the observation of hysteretic magnetic loops with different techniques and different capping layers. On the other hand, the ferromagnetic order of 1 L $Cr_2Ge_2Te_6$ is significantly affected by strong magnetic fluctuations due to its 2D nature. Suppressing the magnetic fluctuations, by introducing a small external magnetic field or an adjacent magnetic pinning layer, may substantially enhance its $T_C$. The realization of macroscopic sized ferromagnetic semiconductor $Cr_2Ge_2Te_6$ films down to 1 L on semiconductor-industry-compatible Si substrate may open a new avenue for spintronic applications based on 2D magnets.

## Methods

**Film growth and characterization.**

The $Cr_2Ge_2Te_6$ film are grown by MBE on Si(111) substrates under a vacuum condition (~ $1\times10^{-8}$ Pa). The Si(111) is flashed . High-purity Cr(99.999%), Ge(99.9999%), Te(99.9999%) are co-evaporated with commercial Knudsen cells. We grow $Cr_2Ge_2Te_6$ layers at 190 °C and anneal at 260 °C for 6h. The growth temperatures for the $Bi_2Te_3$, $Sb_2Te_3$ and CdTe capping layers on $Cr_2Ge_2Te_6$ are 200°C, 200°C, room temperature. The crystal structures are confirmed by x-ray diffraction and atomic force microscopy (Bruker, Innova), and thicknesses of the respective layers are confirmed by STEM.

**Electrical transport measurements.**

The films were patterned into Hall bars. The electrical transport measurements of the Hall bars are performed in a commercial Quantum Design PPMS system. We use the Standard lock-in amplifiers (Stanford Research System SR830) with a low frequency (~15 Hz). The current value for the resistivity and Hall effect measurements was 500 nA.

**SQUID Measurements.**

The samples, approximately 2 mm × 3 mm in size, are encapsulated in a brass tube for out-of-plane measurement and in a quartz tube for in-plane measurement in a Quantum Design MPMS system. We measure the signal of a Si(111) substrate with size of 2 mm × 3 mm under the same conditions. Our final data are obtained by subtracting the signal of the Si(111) substrate and a linear background from the raw data, and no further processing was performed (Fig. S8). Additionally, data spikes resulting from instrument fluctuations are retained in the final data.

**Density functional theory calculations.**

The method employed for determining the Curie temperature was adapted from the approach described in [Nature 546, 265–269 (2017)]. U = 1.2 eV. We define the

magnetic structure as the ABC-stacking hexagonal lattice shown in Fig. S9a. The Hamiltonian is

$$H = \frac{1}{2}\sum_{ll'}\sum_{\nu\nu'} J_{\nu\nu'}^{l-l'} S_{l\nu} \cdot S_{l'\nu'} + \sum_{l}\sum_{\nu} A(S_{l\nu}^z)^2 - g\mu_B \sum_{l}\sum_{\nu} B S_{l\nu}^z$$

where there are N unit cells (indexed by l) and n sublattices in one unit cell (indexed by $\nu$ = 1, 2). Figure S9a and the Hamiltonian are from [Nature 546, 265–269 (2017)].

## Data availability

Relevant data supporting the key findings of this study are available within the article and the Supplementary Information file. All raw data generated during the current study are available from the corresponding authors upon request.

## Acknowledgements

## Author contributions

## Competing interests

The authors declare no competing interests.

## Figure legends

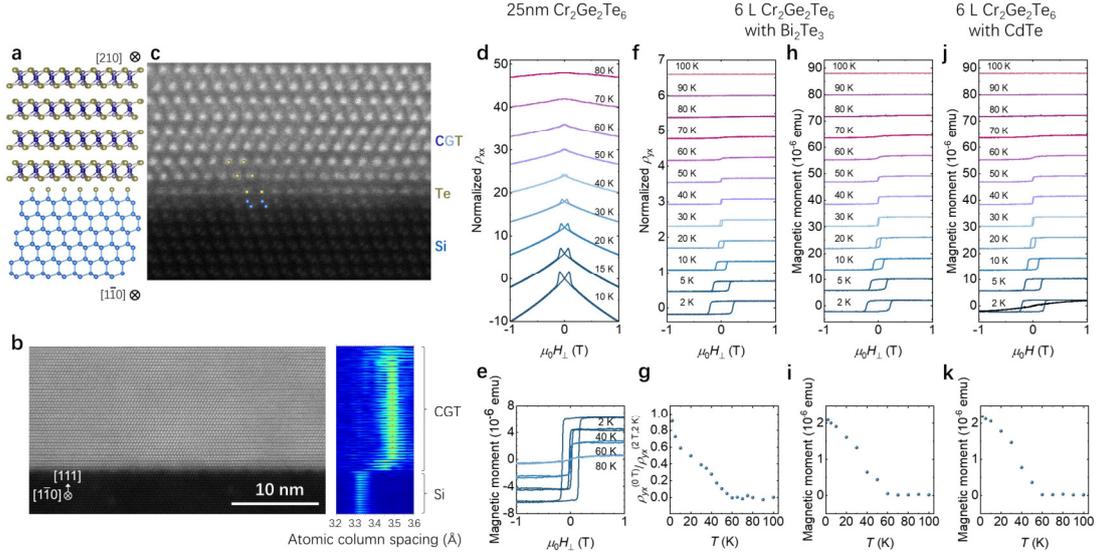

**Fig. 1 | Basic characterization of $Cr_2Ge_2Te_6$ thin films. a** Side view of the $Cr_2Ge_2Te_6$ (top) and the Si(111) (bottom) structure; **b** STEM image of the 25 nm $Cr_2Ge_2Te_6$/Si interface and the Fourier transform (FFT) analysis; **c** Enlarged view of the intermediate interface region in (b), where blue spheres represent Si atoms, and yellow spheres represent Te atoms; **d** Magnetoresistance curves of ~ 25 nm $Cr_2Ge_2Te_6$ at different temperatures, measured under a magnetic field sweep range of ±2 T; **e** Magnetic hysteresis loops of ~ 25 nm $Cr_2Ge_2Te_6$ measured by SQUID at different temperatures; **f** Hall effect measurements of the 6 L $Cr_2Ge_2Te_6$ with a $Bi_2Te_3$ capping layer at different temperatures, with each curve vertically shifted for clarity (field sweep range: ±2 T); **g** Temperature dependence of the zero-field Hall resistance extracted from (g); **h** SQUID-measured out-of-plane magnetic hysteresis loops of the 6 L $Cr_2Ge_2Te_6$ with a $Bi_2Te_3$ capping layer at different temperatures; **i** Temperature dependence of the zero-field magnetic moment extracted from (h); **j** SQUID-measured magnetic hysteresis loops of the 6 L $Cr_2Ge_2Te_6$ with a CdTe capping layer at different temperatures (The black curves at 2 K is applied with the in-plane magnetic field); **k** Temperature dependence of the zero-field magnetic moment extracted from (j).

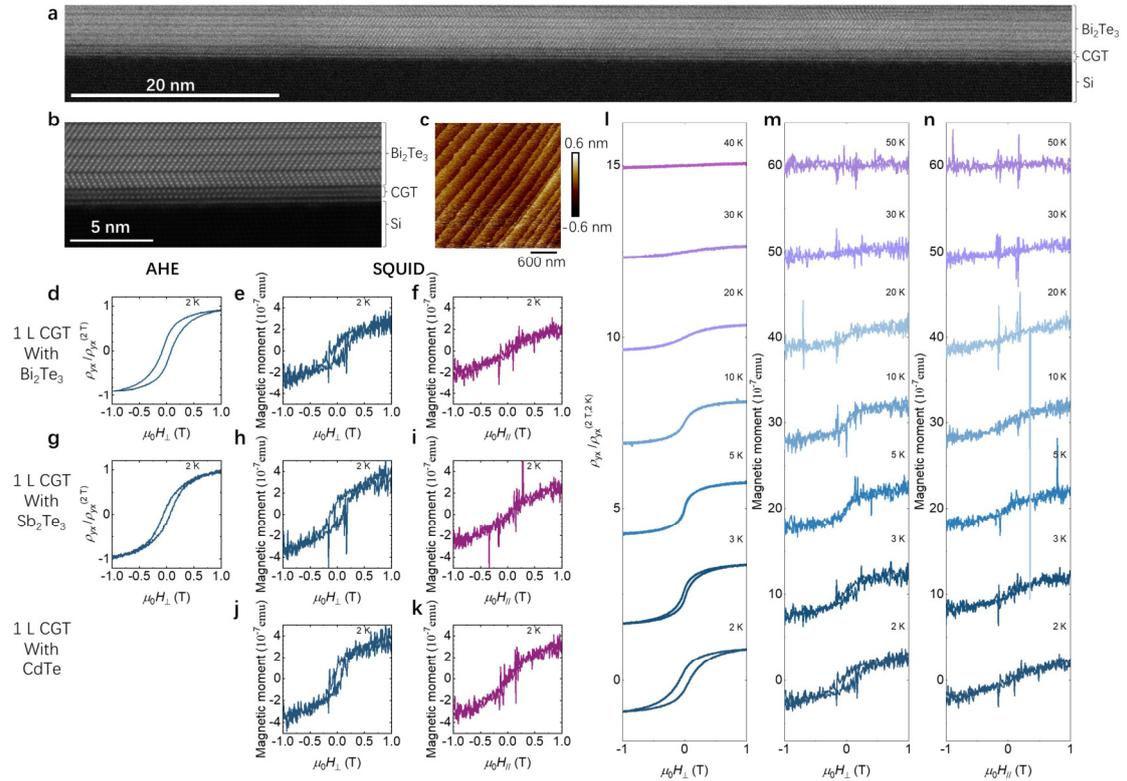

**Fig. 2 | The characterization of 1 L Cr$_2$Ge$_2$Te$_6$ with different capping layers. a-b** STEM image of 1 L Cr$_2$Ge$_2$Te$_6$ with a Bi$_2$Te$_3$ capping layer; **c** The AFM image of 1 L Cr$_2$Ge$_2$Te$_6$; **d** Anomalous Hall effect of 1 L Cr$_2$Ge$_2$Te$_6$ with a Bi$_2$Te$_3$ capping layer under out-of-plane magnetic field at 2 K; **e-f** SQUID-measured magnetic hysteresis loops of 1 L Cr$_2$Ge$_2$Te$_6$ with an Bi$_2$Te$_3$ capping layer under out-of-plane and in-plane magnetic fields at 2 K; **g** Anomalous Hall effect of 1 L Cr$_2$Ge$_2$Te$_6$ with an Sb$_2$Te$_3$ capping layer under out-of-plane magnetic field at 2 K; **h-i** SQUID-measured magnetic hysteresis loops of single-layer Cr$_2$Ge$_2$Te$_6$ with an Sb$_2$Te$_3$ capping layer under out-of-plane and in-plane magnetic fields at 2 K; **j-k** SQUID-measured magnetic hysteresis loops of 1 L Cr$_2$Ge$_2$Te$_6$ with a CdTe capping layer under out-of-plane and in-plane magnetic fields at 2 K; **l** Anomalous Hall effect of 1 L Cr$_2$Ge$_2$Te$_6$ with a Bi$_2$Te$_3$ capping layer at different temperature; **m** SQUID-measured magnetic hysteresis loops of 1 L Cr$_2$Ge$_2$Te$_6$ with a Bi$_2$Te$_3$ capping layer under out-of-plane magnetic field at different temperature; **n** SQUID-measured magnetic hysteresis loops of 1 L Cr$_2$Ge$_2$Te$_6$ with a Bi$_2$Te$_3$ capping layer under in-plane magnetic field at different temperature.

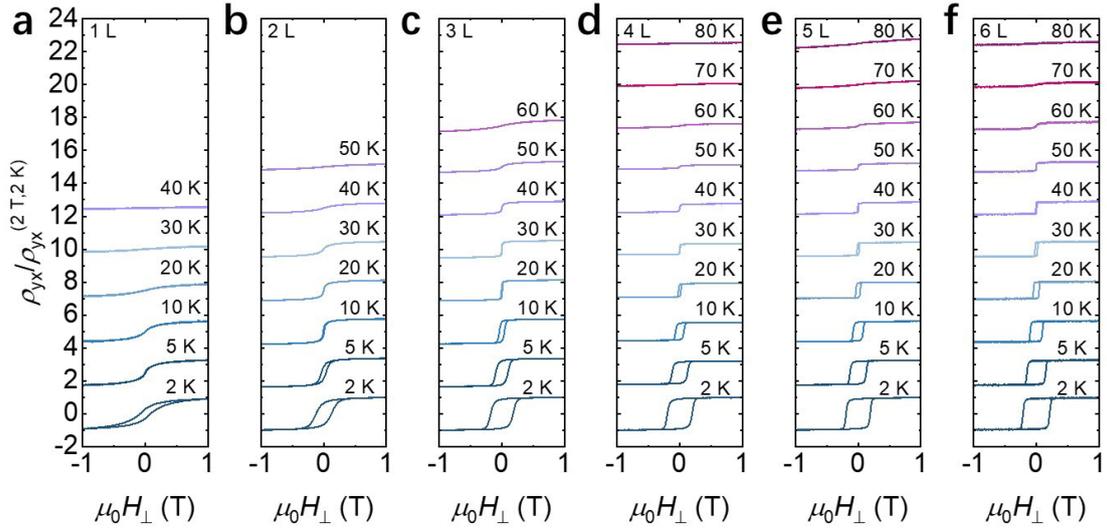

**Fig. 3 | Anomalous Hall effect in 1-6 L Cr$_2$Ge$_2$Te$_6$. a–f** Normalized anomalous Hall effect curves of Cr$_2$Ge$_2$Te$_6$ films with thicknesses of 1 L, 2 L, 3 L, 4 L, 5 L, and 6 L, all covered with a Bi$_2$Te$_3$ capping layer.

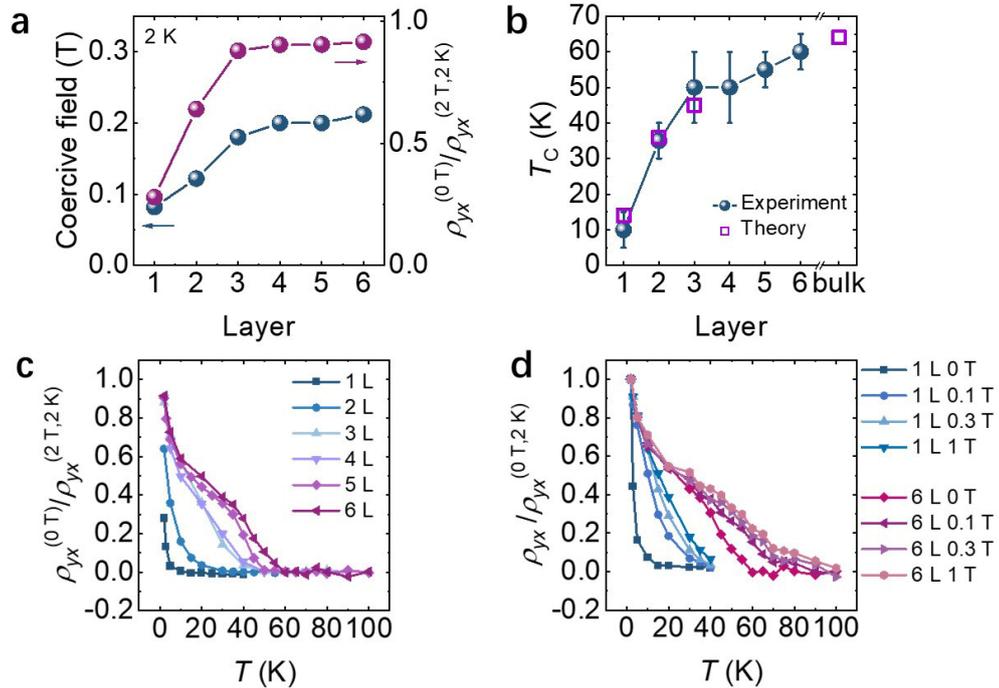

**Fig. 4 | Layer-dependent magnetic ordering in atomically-thin Cr$_2$Ge$_2$Te$_6$. a** Thickness dependence of the coercive field (the blue curve) and remanence ratio ($\rho_{yx}$(0 T)/$\rho_{yx}$(2 T, 2 K)) (the red curve); **b** Thickness dependence of the Curie temperature in experiment (the blue curve) and in theory (the red curve); **c** Zero-field Hall resistance

as a function of temperature for the films in Fig. 3; **d** Temperature dependences of normalized $\rho_{yx}$ of 1 L and 6 L $Cr_2Ge_2Te_6$ at 0 T, 0.1 T, 0.3 T and 1 T.